\documentclass[prd,aps,preprint,preprintnumbers,nofootinbib]{revtex4}
\usepackage{epsfig,footnote}
\usepackage{ulem}
\usepackage{color}
\usepackage{array}
\usepackage{amssymb}
\usepackage{amsmath}
\usepackage{graphicx,subfigure}
\usepackage{longtable}
\usepackage{verbatim}
\usepackage{amsfonts}
\usepackage{hyperref}
\usepackage{cancel}
\newcommand{\degree}{^\circ}
\begin{document}
\title{Analysis of three-body decays $B \to D (V \to ) PP $ under the factorization-assisted
topological-amplitude approach}
\author{Si-Hong Zhou\footnote{shzhou@imu.edu.cn}, Run-Hui Li\footnote{lirh@imu.edu.cn} and Xiao-Yao L\"u}
\affiliation{School of Physical Science and Technology,
Inner Mongolia University, Hohhot 010021, China}
\date{\today}
\begin{abstract}
Motived by the accumulated experimental results on three-body 
charmed $B$ decays with resonance contributions in Babar, LHCb and Belle (II),
we systematically analyze $B_{(s)} \to D_{(s)} (V \to)  P_1 P_2 $ 
decays with $V$ representing a vector resonance ($\rho, K^*, \omega$ 
or $\phi$) and $P_{1,2}$ as a light pseudoscalar meson (pion or kaon). 
The intermediate subprocesses $B_{(s)} \to D_{(s)} V$ are calculated 
with the factorization-assisted topological-amplitude (FAT) approach and 
the intermediate resonant states $V$ described by the relativistic Breit-Wigner 
distribution successively decay to $P_1 P_2$ via strong interaction. 
Taking all lowest resonance states ($\rho, K^*, \omega, \phi$) into account, 
we calculate the branching fractions of these decay modes as well as the 
Breit-Wigner-tail effects for $B_{(s)} \to D_{(s)} (\rho ,\omega \to) KK$. 
Our results agree with the data by Babar, LHCb and Belle (II). 
Among the predictions that are still not observed, there are some 
branching ratios of order $10^{-6}-10^{-4}$ which are hopeful to be 
measured by LHCb and Belle II. Our approach and the 
perturbative QCD approach (PQCD) adopt the compatible theme 
to deal with the resonance contributions. What's more, 
our data for the intermediate two-body charmed $B$-meson decays 
in FAT approach are more precise. As a result, our results for 
branching fractions have smaller uncertainties, especially for 
color-suppressed emission diagram dominated modes.
\end{abstract}

\maketitle

\section{Introduction}\label{Introduction}
Three-body nonleptonic $B$ meson decays not only are important 
for the study of common topics of nonleptonic $B$ meson decays, 
such as testing the standard model (SM), the studying the mechanism 
of $CP$ violation and the emergence of quantum chromodynamics, 
but also provide opportunities for the analysis on the hadron spectroscopy.
Specifically, three-body $B$ meson decays have nontrivial kinematics 
and phase space distributions, which are usually described in terms 
of three two-body invariant mass squared combinations
and two of them constitute two axes to form a Dalitz plot. In the 
edges of the Dalitz plot, the invariant mass squared combinations
of two final-state particles will generally peak as resonances, which 
indicates that intermediate resonances in three-body $B$ meson 
decays show up, and we are able to study the properties of these 
resonances through three-body $B$ meson decays.

The Dalitz plot technique has proven to be a powerful tool to analyze 
the hadron spectroscopy and is widely adopted by experiments. The 
informations on various resonance substructures including the mass, 
spin-parity quantum numbers, etc. have been collected by Babar, 
Belle (II) and LHCb \cite{Belle:2006wbx,BaBar:2009pnd,BaBar:2010rll,
LHCb:2014ioa,LHCb:2015klp,LHCb:2015tsv,LHCb:2018oeb,Belle-II:2023gye}.
Simultaneously, usually under the framework of isobar model
\cite{PhysRevD.11.3165}, these collaborations have also measured 
fit fractions of each resonance and nonresonance components.
 In the isobar model, the total decay amplitude can be expressed 
as a coherent sum of amplitudes of different resonant and nonresonant 
intermediate processes, where the relativistic Breit-Wigner (RBW) model 
usually describe resonant dynamics and exponential distribution for 
nonresonant terms. It is a very good approximation to adopt the RBW 
function for narrow width resonances which can be well separated from 
any other resonant or nonresonant components in the same partial wave,
so that the three-body decays with narrow intermediate states, 
such as $\rho, K^*, \omega, \phi$, have been precisely measured 
by experiments \cite{BaBar:2009pnd,BaBar:2010rll,LHCb:2011yev,
LHCb:2014ioa,BaBar:2015pwa,LHCb:2015klp,LHCb:2015tsv,
LHCb:2018oeb,LHCb:2019sus}.

On the theoretical side, analysis on the nonresonant contributions of 
three-body $B$ meson decays are in an early stage of development.
Approaches or models such as the heavy meson chiral perturbation 
theory \cite{Cheng:2002qu,Cheng:2007si,Cheng:2013dua} and a 
model combing the heavy quark effective theory and chiral Lagrangian
\cite{Fajfer:2004cx} have been applied for calculating the nonresonant 
fraction of three-body charmless $B$ meson decays, such as 
$B \to KKK, \, B \to K \pi \pi$ which are dominated by nonresonant 
contribution \cite{Cheng:2008vy}. More theoretical interest is 
concentrated on the resonant component of three-body $B$ decays,
where two of the three final particles are produced from a resonance
 and recoil against the third meson called a ``bachelor"  meson. 
This type of three-body decay is also called quasi-two-body decay.
Because of the large energy release in $B$ meson decays, the two 
meson pair moves fast antiparrallelly to the bachelor meson in the 
$B$ meson rest frame. Therefore, the interactions between the meson 
pair and the bachelor particle are power suppressed naturally which 
is similar to the statement of ``color transparency" in two-body $B$ 
meson decays. Then approaches based on the factorization hypothesis 
have been proposed for calculating the quasi-two-body decays, such as 
the QCD Factorization (QCDF) 
\cite{Cheng:2002qu, Cheng:2007si, Cheng:2013dua,Huber:2020pqb}, 
the PQCD approach\cite{Chen:2002th,Wang:2014ira,Wang:2016rlo, 
Li:2016tpn, Li:2018psm, Wang:2020plx, Fan:2020gvr, Wang:2020nel, 
Zou:2020atb, Zou:2020fax,Zou:2020ool,Yang:2021zcx,Liu:2021sdw,
Zhang:2022pfn,Zhang:2023uoy,Zhao:2023dnz,Chang:2024qxl} and 
factorization-assisted topological-amplitude (FAT) approach
\cite{Zhou:2021yys,Zhou:2023lbc}.
      
 In this work, we focus on three-body charmed $B$ decays 
 $B_{(s)} \to D_{(s)} P_1 P_2$, where $P_{1,2}$ represents 
 a pion or kaon. Different from charmless decays, intermediate 
 resonances in $B_{(s)} \to D_{(s)} P_1 P_2$ decays are expected 
 to appear in the $m^2 (D P_1)$ and $m^2 (P_1 P_2)$ combinations, 
 thus more resonances, such as charmed states $D^*$ and light 
vector or scalar resonances, can be researched simultaneously 
in one Dalitz plot. In addition, they provide opportunities for 
studies of CP violations. In particular, the Dalitz plot analysis of 
$B^0 \to D K^+ \pi^-$ can be used as a channel to measure the 
unitarity triangle angle $\gamma$ 
\cite{Gershon:2008pe,Gershon:2009qc}
and $B^0 \to \bar D^0 \pi^+ \pi^-$ is sensitive to the $\beta$ angle 
\cite{BaBar:2007fxn,Belle:2006wor}.
 Therefore, much attention has been already paid to 
 $B_{(s)} \to D_{(s)} P_1 P_2$ decays in experiments and 
 theoretical calculations. LHCb Collaborations have investigated 
 structures of ground and excited states of $D^*$, $K^*$ with 
 their corresponding fit fractions through $B_{(s)} \to D_{(s)} K \pi $ 
 \cite{LHCb:2011yev,LHCb:2014ioa}, and $D^*$, $\rho$ in 
 $B_{(s)} \to D_{(s)} \pi  \pi $ by  Babar and Belle 
  \cite{Belle:2006wbx,BaBar:2010rll}.
 Recently, the virtual contribution from $\rho$ in 
 $B_{(s)} \to D_{(s)} KK $ has been measured by Belle II 
 \cite{Belle-II:2023gye}. Motived by the experimental progress on 
 $B_{(s)} \to D_{(s)} P_1 P_2$ decays, theoretical calculation on 
 the branching fractions of various types of charmed quasi-two-body 
 decays, $B_{(s)} \to D_{(s)} \pi \pi $, $B_{(s)} \to D_{(s)} K \pi $ and 
 $B_{(s)} \to D_{(s)} K K $ with intermediate resonances 
 $D^*$, $\rho, K^*, \phi$ have been completed in a series of works 
 with the PQCD approach \cite{Ma:2016csn,Ma:2020dvr,Chai:2021kie, 
 Zou:2022xrr, Fang:2023dcy,Wang:2024enc}.
In FAT approach, we have done a systematic research on 
 $B_{(s)} \to D_{(s)} P_1 P_2$ with ground charmed mesons $D^*$ 
 as the intermediate states and $P_1, P_2$ representing $\pi$ or $ K$  
 \cite{Zhou:2021yys}.  
 The results of $B \to D^*P_2 \to D P_1 P_2$
 in FAT approach are in better agreement with experimental data 
 and more precise than the PQCD approach's predictions.
 
 FAT approach is firstly proposed to resolve the problem about 
 nonfactorizable contributions in two-body $D$ and $B$ meson 
 decays \cite{Li:2012cfa, Li:2013xsa, Zhou:2015jba, Zhou:2016jkv, 
 Jiang:2017zwr, Zhou:2019crd, Zhou:2021yys,Qin:2021tve}  
 and then has been  successfully generalized to quasi-two-body $B$ 
meson decays \cite{Zhou:2021yys,Zhou:2023lbc}. It is based on the 
framework of conventional topological diagram approach, which is 
used to classify the decay amplitudes by different electroweak 
Feynman diagrams, but keeping $SU(3)$ breaking effects. 
Only a few unknown nonfactorization parameters need to be 
fitted globally with all experimental data. Therefore, FAT approach 
is able to provide the most precise decay amplitudes of (intermediate) 
two-body $B$ meson decays especially with charmed meson final
states. Then in a quasi-two-body $B$ meson decay the intermediate 
resonances successively decay to final meson pairs via strong 
interaction, which are described in terms of the usual RBW formalism
as what's done in experiments. Actually, in the PQCD approach the 
light cone distribution amplitude (LCDA) of a meson pair originating 
from a P-wave resonance can be expressed as time-like form 
factors and then is also parameterized by the RBW distribution. 
So their framework of dealing with the resonances is compatible
with that in FAT. More details about this can be found in 
\cite{Wang:2024enc}. Therefore, the main difference between the 
FAT and the PQCD approach in quasi-two-body decays is how to 
calculate the intermediate two-body weak decays. It is well known 
that large nonperturbative contributions and power corrections 
expanded in $m_c/m_b$ of color suppressed and W-exchange 
diagrams in charmed $B$ decays have not been able to be 
calculated in PQCD approach~\cite{Keum:2003js}, which 
results in large uncertainties of PQCD approach's predictions 
for $B \to D P_1 P_2$ with resonances $D^*$, $\rho, K^*, \phi$.
In this paper we will apply FAT approach to study the
$B_{(s)} \to D_{(s)} P_1 P_2$  decay with the ground state 
light vector intermediate resonances $\rho, K^*, \phi,\omega$, 
which are generally the largest components and their fit fractions 
have been well measured separately from any other vector 
resonances in LHCb and Belle II \cite{Belle:2006wbx,LHCb:2014ioa,
LHCb:2015klp,LHCb:2015tsv,LHCb:2018oeb,Belle-II:2023gye}.
 
 This paper is organized as follows. In Sec. \ref{sec:2}, the theoretical 
 framework is introduced. The numerical results and discussions about 
 $B_{(s)} \to D_{(s)} (\rho \to ) \pi \pi $, $B_{(s)} \to D_{(s)} (K^*\to )K \pi $, 
 $B_{(s)} \to D_{(s)} (\phi \to ) K \bar K $ and 
 $B_{(s)} \to D_{(s)} (\rho, \omega \to )K \bar K $ are collected 
 in Sec. \ref{sec:3}. Finally, a summary is given in Section \ref{sec:4} .


\section{FACTORIZATION OF AMPLITUDES FOR TOPOLOGICAL DIAGRAMS}\label{sec:2}

\begin{figure}
\begin{center}
\scalebox{0.6}{\epsfig{file=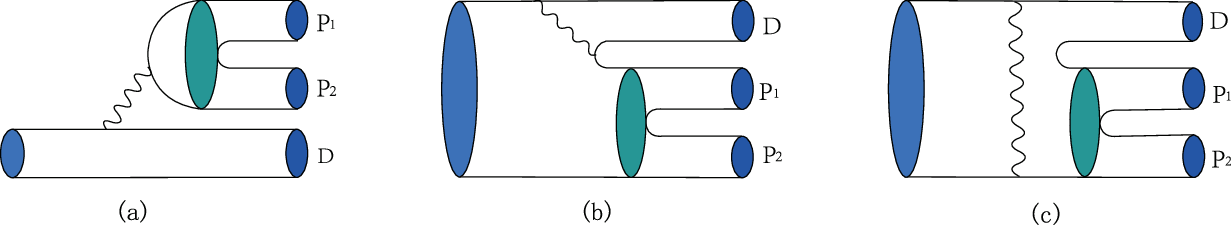}}
\caption{Topological diagrams of 
             $\bar B_{(s)} \to D_{(s)} (V \to ) P_1 P_2$ 
              under the framework of quasi-two-body decay 
              with the wave line representing a $W$ boson:
  (a) the color-favored emission diagram $T$,
  (b) the color-suppressed emission diagram $C$, 
 and (c) the W-exchange diagram $E$.}
\label{TCE}
\end{center}
\end{figure}

\begin{figure}
\begin{center}
\scalebox{0.6}{\epsfig{file=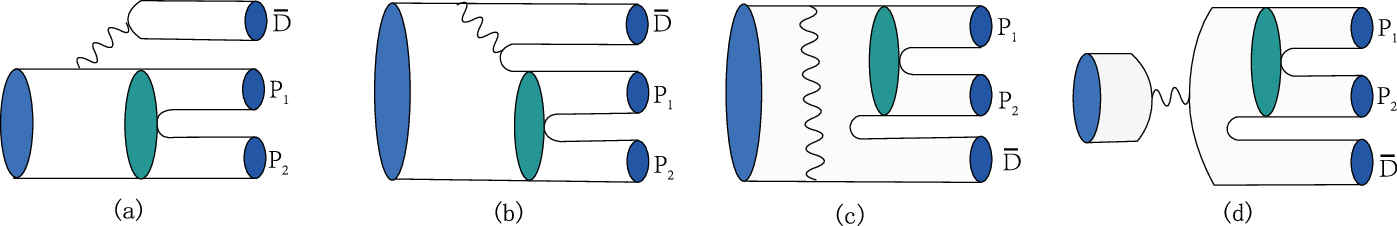}}
\caption{Topological diagrams of 
              $\bar B_{(s)} \to \bar D_{(s)} (V \to ) P_1 P_2$ 
              under the framework of quasi-two-body decay 
              with the wave line representing a $W$ boson:
  (a) the color-favored emission diagram $T$,
  (b) the color-suppressed emission diagram $C$, 
  (c) the W-exchange diagram $E$,
   and (d) W-annihilation diagram $A$.}
\label{TCEA}
\end{center}
\end{figure}

The charmed quasi-two-body decay $B_{(s)} \to D_{(s)} (V \to) P_1 P_2$ 
happens through two subprocesses, where the $D_{(s)}$ meson represents 
$D_{(s)}$ or its antiparticle $\bar D_{(s)}$. $B_{(s)}$ meson decays 
to an intermediate resonant state $D_{(s)} V $ firstly, and subsequently 
the unstable resonance decays to a pair of light psudoscaler, $V \to P_1 P_2$.
The first subprocess at quark level is induced by weak transitions 
$b \rightarrow \,c\, q\, \bar{u} \, \, (q=d, s)$ and 
$b \rightarrow \,u\, q\, \bar{c} \, \, (q=d, s)$ for $D_{(s)}$ and $\bar D_{(s)}$ 
final states, respectively. The secondary one proceeds directly by strong 
interaction. According to the topological structures of 
$b \rightarrow \,c\, q\, \bar{u} $, the diagrams contributing to 
$\bar B_{(s)} \to D_{(s)} (V \to) P_1 P_2$ can be classified 
into three types as listed in Fig.~\ref{TCE},
 a) color-favored emission diagram $T$,
  b) color-suppressed emission diagram $C$, 
  and c) $W$-exchange diagram $E$.
Similarly, besides $T, \, C, \, E$ diagrams, the topologies of 
$\bar B_{(s)} \to \bar D_{(s)} (V \to) P_1 P_2$ induced by 
$b \rightarrow \,u\, q\, \bar{c} $ transitions include an 
additional W-annihilation diagram $A$ as shown in 
Fig.~\ref{TCEA}.

The amplitudes of the first subprocess 
$\bar B_{(s)} \to D_{(s)} V $ can be refered to the ones of 
two-body charmed $B$ decays in FAT approach \cite{Zhou:2015jba}. 
Factorization has been proven in $T$ topology  at high precision. 
However, large nonfactorizable contributions have been found in 
$C$ and $E$. As done in \cite{Zhou:2015jba}, we parameterize 
matrix elements of the nonfactorizable diagrams $C$ and $E$ 
in FAT approach and a well proven factorization formula for $T$, 
which can be expressed as follows, 
\begin{align}\label{bcamp}
\begin{aligned}
T^{D V}&=\sqrt{2}\, G_F\, V_{c b} V_{u q}^* \, a_1(\mu)\,  f_V\,  m_V\, 
F_1^{B \rightarrow D}\left(m_V^2\right)\left(\varepsilon_V^* \cdot p_B\right),
\\
C^{D V}&=\sqrt{2}\, G_F\, V_{c b} V_{u q}^*\,  f_{D_{(s)}}\,  m_V\,  
A_0^{B \rightarrow V}\left(m_D^2\right)\left(\varepsilon_V^* \cdot p_B\right) \, 
\chi^C \, e^{i \phi^C},
\\
E^{D V}&=\sqrt{2}\,  G_F\, V_{c b} V_{u q}^*\, m_V\,  f_B\, 
 \frac{f_{D_{(s)}} f_V}{f_D f_\pi} \, \left(\varepsilon_V^* \cdot p_B\right)\, 
 \chi^E e^{i \phi^E}\, .
\end{aligned}
\end{align}
 So far there is not enough experimental data to do a global fit for 
 $\bar B_{(s)} \to \bar D_{(s)} V$ decays to extract the unknown 
 nonfactorizable parameters in the $C, E$ amplitudes. Therefore, 
 the same nonfactorizable parameters 
 $ \chi^C, \,  \phi^C, \, \chi^E, \,  \phi^E $ as those for 
 $\bar B_{(s)} \to D_{(s)} V$ are adopted in a good approximation,
 just as what we do in Ref.~\cite{Zhou:2015jba}. The topological 
 diagram $A$ for $\bar B_{(s)} \to \bar D_{(s)} V$ is dominated by 
large factorizable contribution and can be calculated in the pole 
model \cite{Zhou:2015jba}, which is given as
\begin{align}\label{buamp}
\begin{aligned}
A^{\bar D V}&=-\sqrt{2}\,  G_F\, V_{u b} V_{c q}^*\,  a_1(\mu)\,   f_B\,  
\frac{ f_{D_{(s)}} g_{DDV} m_{D}^{2} } { m_{B}^{2}-m_{D}^{2}}
(\varepsilon^{*}_{V}\cdot p_{B})\, .
\end{aligned}
\end{align}
For simplification of notations, we omit the subscript `(s)' in $D_{(s)}$ in
eqs. (\ref{bcamp},\ref{buamp}) and the following equations except in 
$f_{D_{(s)}}$. $a_1({\mu})$ is the effective Wilson coefficient for the 
factorizable topologies $T$ and $A$. $\chi^{C(E)}$ and $\phi^{C(E)}$ 
represent the magnitude and associated phase of $C$($E$) diagram 
globally fitted with the experimental data. $f_V$, $f_{D_{(s)}}$ and $f_{B}$ 
are the decay constants of the corresponding vector, $D_{(s)}$ and $B$ 
mesons. $F_1^{B \to D}$ and $ A_0^{B \to V}$ denote the vector form 
factors of $B_{(s)} \to D$ and $B_{(s)} \to V$ transitions, which depend 
on the square of transfer momentum $Q^2$ and can be parameterized 
in pole model as,
\begin{equation}\label{eq:ffdipole}
F_{i}(Q^{2})={F_{i}(0)\over 1-\alpha_{1}{Q^{2}\over m_{\rm pole}^{2}}
+\alpha_{2}{Q^{4}\over m_{\rm pole}^{4}}},
\end{equation}
where $F_{i}$ represents $F_{1}$ or $A_{0}$, and $m_{\rm pole}$ 
is the mass of the corresponding pole state, such as $B^{*}$ for 
$F_{1}$ and $B$ for $A_{0}$. $\alpha_{1,2}$ are the model parameters.
 In eq.(\ref{buamp}), $g_{DDV}$ is the effective strong coupling constant 
 and its value can be obtained from the vector meson dominance 
 model \cite{PhysRevC.62.034903}, $g_{DDV}=2.52$.
 
 Next we will illustrate the calculation of the second subprocess, 
 e.g. intermediate resonances decay to final states via strong interaction, 
 $V \to P_1 P_2$. As what's done in experiments  
 \cite{BaBar:2011vfx, BaBar:2012bdw, LHCb:2019sus}, we also adopt 
 the RBW distribution for $\rho$,\, $K^*$,\, $\omega$, and $\phi$ 
 resonances, which is expressed as \cite{Cheng:2013dua},
 \begin{align}\label{RBW}
L^{\mathrm{RBW}}(s)=\frac{1}{s-m_{V}^{2}+i m_{V} \Gamma_{V}(s)}\, ,
\end{align}
where $s$ represents the invariant mass square of meson pair 
with 4-momenta $p_1,\,  p_2$, $s=(p_1+p_2)^2$. 
$\Gamma_{V}(s)$ represents $s$-dependent width of vector resonances 
and is defined as 
\begin{align}\label{width}
\Gamma_{V}(s)=\Gamma_{0}\left(\frac{q}{q_{0}}\right)^{3}
\left(\frac{m_{V}}{\sqrt{s}}\right) X^{2}\left(q\, r_{\mathrm{BW}}\right)\, ,
\end{align}
where $X(q\, r_{\mathrm{BW}})$ is the Blatt-Weisskopf barrier factor,
\begin{equation}
X\left(q\, r_{\mathrm{BW}}\right)
=\sqrt{[1+\left(q_{0}\, r_{\mathrm{BW}}\right)^{2}]/[1+
\left(q\, r_{\mathrm{BW}}\right)^{2}}]\, .
\end{equation}
In the above two equations, 
$q=\frac{1}{2} \sqrt{\left[s-\left(m_{P_1}+m_{P_2}\right)^{2}\right]
 \left[s-\left(m_{P_1}-m_{P_2}\right)^{2}\right] / s}$ is momentum 
 magnitude of the final state $P_1$ or $P_2$ in the rest frame 
of resonance $V$, and $q_0$ is just value of $q$ when intermediate 
resonance is on-shell, $s = m^2_{V}$. While for the case that the 
pole mass locates outside the kinematics region, 
i.e., $m_{V}<m_{P_1}+m_{P_2}$, $m_V$ needs to be replaced by 
an effective mass $m^{\mathrm{eff}}_{V}$ so that 
$\sqrt s = m^{\mathrm{eff} }_{V}$ for $q_0$. The effective mass 
$m^{\mathrm{eff}}_{V}$ is given by the ad hoc formula
 \cite{Aaij:2014baa, Aaij:2016fma},
\begin{align}\label{massDstar}
 m_{V}^{\text {eff }}\left(m_{V}\right)
 =m^{\min }+\left(m^{\max }-m^{\min }\right)
 \left[1+\tanh \left(\frac{m_{V}-\frac{m^{\min }+
 m^{\max }}{2}}{m^{\max }-m^{\min }}\right)\right]\, ,
 \end{align}
 where $m^{\max }$ ($m^{\min }$) is the upper (lower) boundary 
 of the kinematics region. Another parameter together with $q$ in 
 $X (q\, r_{\mathrm{BW}})$ is the the barrier radius with its value 
 $r_{\mathrm{BW}}=4.0 (\mathrm{GeV})^{-1}$ for all resonances 
 \cite{LHCb:2019sus}. $\Gamma_{0}$ in eq.(\ref{width}) represents 
 the full widths of the resonant states and their values are taken from 
 Particle Data Group (PDG)~\cite{Workman:2022ynf} and listed in 
 Table \ref{tab:mass and width} together with their masses $m_V$.
\begin{table}[tbhp]
\caption{Masses $m_V$ and full widths $\Gamma_0$   
of vector resonant states.\cite{Workman:2022ynf} }
\vspace{3mm}
\label{tab:mass and width}
\centering
\begin{tabular}{cccc}
\hline
Resonance ~~&~~ Line shape Parameters~~ &
~~Resonance ~~&~~ Line shape Parameters
\\ \hline
$\rho(770)~~$  & $m_V\, =\, 775.26\,  \, \mathrm{MeV}$ &
$\omega(782)~~$  & $m_V\, =\, 782.65\,  \, \mathrm{MeV}$ \\
& $\Gamma_0\, =\, 149.1\, \, \mathrm{MeV}$
& &$\Gamma_0\, =\, 8.49\, \, \mathrm{MeV}$\\
$K^*(892)^+~~$  & $m_V\, =\, 891.66\,  \, \mathrm{MeV}$ 
&$K^*(892)^0~~$  & $m_V\, =\, 895.55\,  \, \mathrm{MeV}$\\
& $\Gamma_0\, =\, 50.8\, \, \mathrm{MeV}$
 & &$\Gamma_0\, =\, 47.3\, \, \mathrm{MeV}$\\
$\phi(1020)~~$ & $m_V\, =\, 1019.46\,  \, \mathrm{MeV}$ \\
&  $\Gamma_0\, =\, 4.25 \, \, \mathrm{MeV}$\\
\hline
\end{tabular}
\end{table}

After getting the distribution function of the vector resonances, 
we proceed to consider matrix element 
$\left\langle  P_2 \left(p_{2}\right) P_3 \left(p_{3}\right) | V (p_V)\right\rangle$.
It can be parametrized as a strong coupling constant $g_{ V P_1 P_2}$ 
which describes the strong interactions of the three mesons at hadron level.
Inversely, the strong coupling constant $g_{V P_1 P_2}$ can be extracted 
from the partial decay widths $\Gamma_{V  \to P_1 P_2}$ by
\begin{align}
\Gamma_{V \rightarrow P_{1} P_{2}}
=\frac{2}{3} \frac{p_{c}^{3}}{4 \pi m_{V}^{2}} g_{V P_{1} P_{2}}^{2}\, ,
\end{align}
where $p_c$ is the magnitude of one pseudoscalar meson's momentum in 
the rest frame of the mother vector meson. The numerical results of
 $g_{\rho \rightarrow \pi^{+} \pi^{-}}$, $g_{K^{*} \rightarrow K^{+} \pi^{-}}$, 
and $g_{\phi \rightarrow K^{+} K^{-}}$ have already been directly extracted 
from experimental data~\cite{Cheng:2013dua},
\begin{align}\label{gVPP}
&g_{\rho \rightarrow \pi^{+} \pi^{-}}=6.0\, ,
\quad g_{K^{*} \rightarrow K^{+} \pi^{-}}=4.59\, ,
\quad g_{\phi \rightarrow K^{+} K^{-}}=-4.54\, .
\end{align}
Those strong coupling constants, which can not be extracted directly 
with experimental data, can be related to the ones in Eq.(\ref{gVPP} ) 
by employing the quark model\cite{Bruch:2004py},
$$g_{\rho \rightarrow K^{+} K^{-}}: g_{\omega \rightarrow K^{+} K^{-}}: 
g_{\phi \rightarrow K^{+} K^{-}}=1: 1:- \sqrt{2},$$
$$g_{\rho^{0} \pi^{+} \pi^{-}} = g_{\rho^+ \pi^0 \pi^+}\, , 
\, g_{\rho^{0} \pi^{0} \pi^{0}}= g_{\omega \pi^+ \pi^-} =0  \, ,$$
$$g_{\rho^{0} K^{+} K^{-}}=-g_{\rho^{0} K^{0} \bar{K}^{0}}=
g_{\omega K^{+} K^{-}}=g_{\omega K^{0} \bar{K}^{0}}, \, 
g_{\phi K^{+} K^{-}}=g_{\phi K^{0} \bar{K}^{0}}\, .$$

Finally, combing the two subprocesses together, one can get 
the decay amplitudes of the topological diagrams for 
$B \to D (V \to) P_1 P_2$ shown in Fig.\ref{TCE} and 
Fig.\ref{TCEA}, which are given as 
\begin{align}\label{TCEamp}
\begin{aligned}
T&=
\left\langle P_1 \left(p_{1}\right) P_2 \left(p_{2}\right) |(\bar q u)_{V-A} |0 \right \rangle
\left\langle D (p_D)|(\bar{c} b)_{V-A}| B(p_B) \right \rangle \\
&=\frac{\left\langle  P_1 \left(p_{1}\right) P_2 \left(p_{2}\right)| V(p_V)\right\rangle}
{s-m_{V}^{2}+i m_{V} \Gamma_{V}(s)}
\left\langle V (p_V) \left|(\bar{q} u)_{V-A}\right| 0 \right\rangle
\left \langle D \left(p_{D}\right)\left|(\bar{c} b)_{V-A}\right| B(p_B)\right\rangle\\
&=p_D \cdot\left(p_{1}-p_{2}\right)\, \sqrt 2\, G_{F}\, V_{c b} V_{u q}^{*}\, 
a_1\, f_{V} m_{V} F_1^{B \to D} (s)\, 
\frac{g_{V P_1 P_2}} {s-m_{V}^{2}+i m_{V} \Gamma_{V}(s)} \, ,
 \\
C&=
\left\langle P_1 \left(p_{1}\right) P_2 \left(p_{2}\right)\left|(\bar{q} b)_{V-A}
\right| B (p_B) \right\rangle\left \langle D \left(p_{D}\right)
\left|(\bar{c} u)_{V-A}\right| 0\right\rangle\\
&=\frac{\left\langle  P_1 \left(p_{1}\right) P_2 \left(p_{2}\right) | V (p_V)\right\rangle}
{s-m_{V}^{2}+i m_{V} \Gamma_{V}(s)}
\left\langle V (p_V) \left|(\bar{q} b)_{V-A}\right| B(p_B) \right\rangle
\left \langle D \left(p_{D}\right)\left|(\bar{c} u)_{V-A}\right| 0\right\rangle\\
&=p_D \cdot\left(p_{1}-p_{2}\right)\, \sqrt 2\, G_{F}\, V_{c b} V_{u q}^{*} \,
 f_{D}\, m_{V} A_0^{B \to V} (m_{D}^2)\, \chi^C \mathrm{e}^{\phi^C}\,
  \frac{g_{V P_1 P_2}} {s-m_{V}^{2}+i m_{V} \Gamma_{V}(s)} \,  ,
 \\
 E&=\left\langle D \left(p_{D}\right) P_1 \left(p_{1}\right)  P_2 (p_2) 
 \left|\mathcal{H}_{eff}\right| B(p_B) \right\rangle
\\
&=\frac{\left\langle  P_1 \left(p_{1}\right) P_2 \left(p_{2}\right)| V(p_V)\right\rangle}
{s-m_{V}^{2}+i m_{V} \Gamma_{V}(s)}
\left\langle D(p_D) V(p_V)  \left|\mathcal{H}_{eff} \right| B(p_B)\right\rangle\\
&=p_D \cdot\left(p_{1}-p_{2}\right)\, \sqrt 2\, G_{F} \, V_{c b} V_{u q}^{*}\, 
 m_{V}\, f_{B}\,  \frac{f_V f_{D_{(s)}}}{f_{\pi} f_{D}}\,  
 \chi^{E} \mathrm{e}^{i \phi^{E}} \, 
\frac{g_{V P_1 P_2}} {s-m_{V}^{2}+i m_{V} \Gamma_{V}(s)} \, ,
\end{aligned}
\end{align}
for $b \to c$ transition, and
\begin{align}\label{TCEAamp}
\begin{aligned}
T&=
\left\langle P_1 \left(p_{1}\right) P_2 \left(p_{2}\right) |(\bar u b)_{V-A} |B(p_B) \right \rangle
\left\langle \bar D (p_{\bar D})|(\bar{c} q)_{V-A}| 0 \right \rangle \\
&=\frac{\left\langle  P_1 \left(p_{1}\right) P_2 \left(p_{2}\right)| V(p_V)\right\rangle}
{s-m_{V}^{2}+i m_{V} \Gamma_{V}(s)}
\left\langle V (p_V) \left|(\bar{u} b)_{V-A}\right| B(p_B) \right\rangle
\left \langle \bar D \left(p_{\bar D}\right)\left|(\bar{c} u)_{V-A}\right| 0\right\rangle\\
&=p_{\bar D} \cdot\left(p_{1}-p_{2}\right)\, \sqrt 2 \, G_{F}\, V_{u b} V_{c q}^{*}\, a_1\,
 f_{D} \, m_{V} \, A_0^{B \to V} (m_D^2)\, 
\frac{g_{V P_1 P_2}} {s-m_{V}^{2}+i m_{V} \Gamma_{V}(s)} \, ,
\\
C&=
\left\langle P_1 \left(p_{1}\right) P_2 \left(p_{2}\right)
\left|(\bar{u} b)_{V-A}\right| B (p_B) \right\rangle
\left \langle \bar D \left(p_{\bar D}\right)\left|(\bar{c} q)_{V-A}\right| 0\right\rangle\\
&=\frac{\left\langle  P_1 \left(p_{1}\right) P_2 \left(p_{2}\right) | V (p_V)\right\rangle}
{s-m_{V}^{2}+i m_{V} \Gamma_{V}(s)}
\left\langle V (p_V) \left|(\bar{q} b)_{V-A}\right| B(p_B) \right\rangle
\left \langle \bar D \left(p_{\bar D}\right)\left|(\bar{c} u)_{V-A}\right| 0\right\rangle\\
&=p_{\bar D} \cdot\left(p_{1}-p_{2}\right)\, \sqrt 2\,  G_{F}\,  V_{u b} V_{c q}^{*} \, 
f_{D}\, m_{V}\,  A_0^{B \to V} (m_{D}^2)\,
 \chi^C \mathrm{e}^{i \phi^C}\, \frac{g_{V P_1 P_2}} 
 {s-m_{V}^{2}+i m_{V} \Gamma_{V}(s)} \,  ,
 \\
 E&=\left\langle \bar D \left(p_{\bar D}\right) 
 P_1 \left(p_{1}\right)  P_2 (p_2) \left|\mathcal{H}_{eff}\right| B(p_B) \right\rangle
\\
&=\frac{\left\langle  P_1 \left(p_{1}\right) P_2 \left(p_{2}\right)| V(p_V)\right\rangle}
{s-m_{V}^{2}+i m_{V} \Gamma_{V}(s)}
\left\langle \bar D(p_{\bar D}) V(p_V)  \left|\mathcal{H}_{eff} \right| B(p_B)\right\rangle\\
&=p_{\bar D} \cdot\left(p_{1}-p_{2}\right)\, \sqrt 2 \, G_{F}\, V_{u b} V_{c q}^{*}\, 
 m_{V}\,  f_{B} \, \frac{f_V f_{D_{(s)}}}{f_{\pi} f_{D}}\,  \chi^{E} \mathrm{e}^{i \phi^{E}} \, 
\frac{g_{V P_1 P_2}} {s-m_{V}^{2}+i m_{V} \Gamma_{V}(s)} \, ,
\\
A&=\left\langle \bar D \left(p_{\bar D}\right) P_1 \left(p_{1}\right)  P_2 (p_2)
 \left|\mathcal{H}_{eff}\right| B(p_B) \right\rangle
\\
&=\frac{\left\langle  P_1 \left(p_{1}\right) P_2 \left(p_{2}\right)| V(p_V)\right\rangle}
{s-m_{V}^{2}+i m_{V} \Gamma_{V}(s)}
\left\langle \bar D(p_{\bar D}) V(p_V)  \left|\mathcal{H}_{eff} \right| B(p_B)\right\rangle\\
&= p_{\bar D} \cdot\left(p_{1}-p_{2}\right)\, \sqrt{2}\,  G_F\, V_{u b} V_{c q}^*\,  a_1\,  
 f_B\,  { f_{D} g_{DDV} m_{D}^{2}\over m_{B}^{2}-m_{D}^{2}}\,
  \frac{g_{V P_1 P_2}} {s-m_{V}^{2}+i m_{V} \Gamma_{V}(s)} \, ,
\end{aligned}
\end{align}
for $b \to u$ transition, respectively.
In the above equations $q=d, s$ and $p_V=p_1+p_2=\sqrt{s}$.
The decay amplitudes of $B  \to  D P_1 P_2$ in 
eq.(\ref{TCEamp}) and eq.(\ref{TCEAamp}) 
can also be formally written as 
\begin{align}
\left\langle D \left(p_{D}\right) P_1 \left(p_{1}\right)  P_2 (p_2) 
 \left| \mathcal{H}_{eff} \right| B (p_B) \right\rangle
= p_D \cdot\left(p_{1}-p_{2}\right) \mathcal{A}(s)\, ,
\end{align}
where $\mathcal{A}(s)$ represents the sub-amplitudes in 
Eqs.(\ref{TCEamp}-\ref{TCEAamp}) with the factor 
$p_D \cdot\left(p_{1}-p_{2}\right)$ taken out.
The differential width of $B \to D P_1 P_2$ is
\begin{align}\label{dwidth}
\begin{aligned}
d \, \Gamma
=&d\, s  \,  \frac{1}{(2 \pi)^3}\,
\frac{(\left|\mathbf{p_D} \| \mathbf{p}_{1}\right|\,)^3}{6 m_B^3}\, | \mathcal{A}(s)|^2\, , \\
\end{aligned}
\end{align}
where $|\mathbf{ p_{D}} |$ and $|\mathbf{ p_{1}} |$ represent the 
magnitudes of the momentum $p_D$ and $p_1$, respectively. 
In the rest frame of the vector $V$ resonance, their expressions are
\begin{align}
|\mathbf{ p_{D}} |=&\frac{1}{2\, \sqrt s}\,
\sqrt{ (m_B^2-m_{D}^2)^2  \, -2(m_B^2+m_{D}^2) \,s +s^2}\, \nonumber\\
|\mathbf{ p_{1}} |=&\frac{1}{2\, \sqrt s}\,\sqrt{ \left[s-(m_{P_1}+m_{P_2})^2 \right] \,
 \left[s-(m_{P_1}-m_{P_2})^2 \right]}\, ,
\end{align}
where $|\mathbf{ p_{1}} |=q$.
\section{Numerical results and discussion}\label{sec:3}
The input parameters are classified into 
(a) electroweak coefficients: Cabibbo-Kobayashi-Maskawa (CKM) 
matrix elements and Wilson coefficients; 
(b) nonperturbative QCD parameters: decay constants, transition form factors 
and nonfactorizable parameters $\chi^{C(E)}$, $\phi^{C(E)}$; 
(c) Hadronic parameters: $m_V$, $\Gamma_0$, and $g_{V P_1 P_2}$ 
involved in strong interaction decays of vector mesons, which have been 
listed in tab.~\ref {tab:mass and width} given in previous section.
The Wolfenstein parametrization of the CKM matrix is utilized with the
 Wolfenstein parameters as~\cite{Workman:2022ynf}
$$ \lambda=0.22650 \pm 0.00048,~~~A=0.790^{+0.017}_{-0.012},
~~~\bar \rho=0.141^{+0.016}_{-0.017},~~~\bar \eta=0.357 \pm 0.01.$$
 The decay constants of pseudoscalar mesons and light vector mesons,
 and transition form factors of $B$ meson decays $F_1(Q^2)$ and $A_0(Q^2)$ 
 at recoil momentum square $Q^2=0$ are listed in Tables \ref{tab:decay constants} 
 and \ref{tab:ff}, respectively. The decay constants of $\pi, K,D$ and $B$ 
are from the PDG by global fit with experimental data \cite{Workman:2022ynf}.
The remaining input nonperturbative QCD parameters, the decay constants of 
$D_s$ and $B_s$, and all form factors are obtained from various theoretical results, 
such as light-cone sum rules \cite{Wang:2015vgv,Gao:2019lta,Cui:2022zwm}.
We will utilize the same theoretical values as in the previous work 
by two of us (S.-H. Z. and C.-D. L.) with other colleagues~\cite{Zhou:2015jba}, 
with 5\% uncertainty kept for decay constants and 10\% uncertainty for 
form factors. Also as done in \cite{Zhou:2015jba}, the dipole model of 
form factors is adopted with dipole parameters $\alpha_{1,2}$ listed in 
Tab. \ref{tab:ff}.
\begin{table}
\caption{The decay constants of light pseudoscalar mesons and vector mesons 
 (in units of MeV). }
\vspace{3mm}
\label{tab:decay constants}
\newsavebox{\tablebox}
\begin{lrbox}{\tablebox}
\centering
\begin{tabular}{ccccccccccccc|}
\hline
$f_{\pi}$ & $f_{K}$  & $f_{D}$ & $f_{D_s}$ &  $f_{B}$ & $f_{B_s}$ & 
$f_{\rho}$ & $f_{K^*}$ & $f_{\omega}$ & $f_{\phi}$ &
\\\hline
$130.2 \pm 1.7$ & $155.6 \pm 0.4$ &$211.9 \pm 1.1$&$258 \pm 12.5 $& 
$190.9 \pm 4.1$ & $225 \pm 11.2$ & $213\pm11$ &$220\pm11$& $192\pm10$
& $225\pm11$
\\
\hline
\end{tabular}
\end{lrbox}
 \scalebox{0.9}{\usebox{\tablebox}}
\end{table}
\begin{table} [hbt]
\caption{Transition form factors at $q^2=0$ and dipole model parameters
 used in this work. }\label{tab:ff}
\vspace{3mm}
\centering
\begin{tabular}{|c|c|c|c|c|c|c|c|c|c|c|c|c|c|}
\hline&
$~~F_{1}^{B\to D}~~$&
$~~F_{1}^{B_s\to D_s}~~$&
$~~A_{0}^{B\to \rho}~~$&
$~~A_{0}^{B \to K^*}~~$&
$~~A_{0}^{B_s \to K^*}~~$&
$~~A_{0}^{B \to \omega}~~$&
$~~A_{0}^{B_s \to \phi}~~$\\
\hline
$~F_i(0)~$&
0.54&
0.58&
0.30&
0.33&
0.27&
0.26&
0.30\\
$\alpha_1$&
2.44&
2.44&
1.73&
1.51&
1.74&
1.60&
1.73\\
$\alpha_2$&
1.49&
1.70&
0.17&
0.14&
0.47 &
0.22 &
0.41\\
\hline
\end{tabular}
\end{table}
The effective Wilson coefficients $a_1$ is $1.036$ calculated 
at scale $\mu=m_b/2$. The nonfactorizable parameters
 $\chi^{C(E)}$ and $\phi^{C(E)}$ fitted with experimental 
 data \cite{Zhou:2015jba} are 
\begin{align}
\begin{aligned}
\label{fitpara}
\chi^{C}&=0.48 \pm 0.01, ~~~~~ 
\phi^{C}=\left(56.6_{-3.8}^{+3.2}\right)^{\degree}, \\
 \chi^{E}&=0.024_{-0.001}^{+0.002},~~~~~ 
 \phi^{E}=\left(123.9_{-2.2}^{+3.3}\right)^{\degree}.
 \end{aligned}
\end{align}

With all the inputs, we integrate the the differential width 
in Eq.(\ref{dwidth}) over the kinematics region to obtain 
the branching fractions of $\bar B \to  D (V \to) P_1 P_2 $ 
and $ \bar B\to  \bar D (V \to) P_1 P_2 $. Specifically, 
the numerical results for $\bar B_{(s)} \to D (\rho\to) \pi \pi$, 
$\bar B_{(s)} \to D (K^*\to) K \pi$, $\bar B_{(s)} \to D (\phi \to) KK $ 
and $\bar B_{(s)} \to D (\rho, \omega\to) K K$, together with their 
corresponding doubly CKM suppressed decays 
$ \bar B\to  \bar D (V \to) P_1 P_2 $, are collected in Tables 
\ref{Drho},  \ref{DKstar}, \ref{Dphi} and \ref{Domega}, respectively.
In our results denoted by ${\mathcal B}_{\rm {FAT}} $, the uncertainties 
are in sequence from the fitted parameters, form factors, decay 
constants for $ \bar B\to D (V \to) P_1 P_2 $, and an additional 
error from $V_{ub}$ for $\bar B\to  \bar D (V \to) P_1 P_2 $ decays 
induced by $b \to u $ transitions. One can see that the dominating
 errors are from the uncertainties of form factors, which can be improved by
  more precise calculations. Besides the CKM matrix elements 
shown in these tables we also list the intermediate resonance 
decays as well as the topological contributions $T, C, E$ and $A$ 
for convenience of analyzing hierarchies of branching fractions in 
the following. Experimental data in the third column and the 
results in PQCD approach in last column are also list for comparison.

\begin{table}[tbhp]
\caption{Branching ratios in FAT approach of  quasi-two-body decays (top)
$\bar B_{(s)} \to D (\rho\to) \pi \pi$ ($\times 10^{-4}$), 
and (bottom) $\bar B_{(s)} \to \bar D (\rho \to) \pi \pi$ ($\times 10^{-6}$),
 together with results in PQCD and experimental data.
The CKM matrix elements and characters $T$, $C$, $E$ and $A$ 
representing the corresponding topological diagram 
contributions are also listed in the second column.}
 \label{Drho}
\begin{center}
\begin{tabular}{ccccc}
 \hline \hline
Decay Modes &  Amplitudes & Data & ${\mathcal B}_{\rm {FAT}} $  
& ${\mathcal B}_{\rm {PQCD}} $ 
   \\
\hline
   $\bar B \to D(\rho \to) \pi \pi $ & $V_{cb} V_{ud}^*$ &  &  &  \\
   
  $B^- \to D^0 (\rho^-\to)\pi^0 \pi^-$ & $T+C$ & $134 \pm 18 $ &
          $97.7^{+2.1+16.8+8.5}_{-2.3-15.8-8.1}$ & $~~115^{+59}_{-38}~~$
          \\

  $\bar {B}^0 \to D^+ (\rho^-\to)\pi^0 \pi^-$ & $T+E$ & $76 \pm 12$ &  
          $60.0^{+0.5+13.0+6.4}_{-0.3-11.7-6.0}$ & $82.3^{+49.2}_{-29.0}$
          \\  
                             
 $\bar{B}^0 \to D^0(\rho^0\to)\pi^+ \pi^-$ & $\frac{1}{\sqrt 2}(E-C) $ & $3.21 \pm 0.21$ &
          $ 2.50^{+0.14+0.24+0.03}_{-0.13-0.49-0.03} $ & $1.39^{+1.24}_{-0.90}$
          \\
  
  $\bar{B}_s^0 \to D_s^+(\rho^-\to)\pi^- \pi^0$ & $T$  &$95 \pm 20$ &
              $74.5 ^{+0.0+15.6+7.9}_{-0.0-14.2-7.5}$ & $77.2^{+40.2}_{-25.6}$
              \\      
              
      & $V_{cb} V_{us}^*$  & & & \\
  $\bar{B}_s^0 \to D^+(\rho^-\to) \pi^- \pi^0$ & $E$ & &
           $0.018^{+0.003+0+0.005}_{-0.001-0-0.004} $ & $0.051^{+0.022}_{-0.014} $ \\       
                                     
   $\bar{B}_s^0 \to D^0 (\rho^0\to) \pi^+ \pi^-$ & $\frac{1}{\sqrt 2} E$ & &
           $0.009^{+0.002+0+0.002}_{-0.001-0-0.001}$ & $0.026^{+0.010}_{-0.006} $ \\                   
 \hline
 
  $\bar B \to \bar D (\rho \to) \pi \pi $ & $V_{ub} V_{cs}^*$ &  && \\    
  
  $ B^{-} \to D_{s}^{-} (\rho^{0}\to) \pi^{+}\pi^{-}$ & $\frac{1}{\sqrt{2}}T$ &&
             $16.7^{+0.0+3.5+1.7+1.5}_{-0.0-3.2-1.6-1.5}$ & $15.2^{+11.1}_{-8.2} $  \\
             
  $\bar{B}^{0}\to D_{s}^{-} (\rho^{+} \to) \pi^{+}\pi^{0}$ & $T$ &&
            $29.7 ^{+0.0+6.2+2.9+2.6}_{-0.0-5.6-2.8-2.6}$ & $28.2^{+20.4}_{-15.3} $  \\
     
  $\bar{B}_{s}^{0} \to D^{-} (\rho^{+}\to ) \pi^{+}\pi^{0}$ & $E$ &&
           $0.19^{+0.03+0+0.02+0.02}_{-0.02-0-0.02-0.02}$ & $0.69^{+0.20}_{-0.16}$ \\
           
  $\bar{B}_{s}^{0}\to \bar{D}^{0} (\rho^{0} \to) \pi^{+}\pi^{-}$ & $\frac{1}{\sqrt{2}}E$ &&
          $0.09^{+0.02+0+0.01+0.01}_{-0.01-0-0.01-0.01} $  & $0.34^{+0.10}_{-0.08}$  \\
                 
  & $V_{ub} V_{cd}^*$ &  &  &\\      
        
  $ B^{-}\to D^{-} (\rho^{0}\to) \pi^{+}\pi^{-}$ & $\frac{1}{\sqrt{2}}(T-A)$ &&
          $0.35 ^{+0+0.10+0.01+0.03}_{-0-0.09-0.01-0.03}$   &  $0.53^{+0.36}_{-0.27}$    \\
          
  $ B^{-}\to \bar{D}^{0} (\rho^{-} \to) \pi^{+}\pi^{0}$ & $C+A$ &&
         $0.48^{+0.02+0.07+0.01+0.04}_{-0.02-0.06-0.01-0.04} $  &  $0.05^{+0.02}_{-0.01}$\\

  $\bar{B}^{0} \to D^{-} (\rho^{+}\to) \pi^{+}\pi^{0}$ & $T+E$ &&
          $1.03 ^{+0.01+0.23+0.01+0.09}_{-0.01-0.20-0.01-0.09}$  & $0.76^{+0.59}_{-0.31}$ \\
          
  $\bar{B}^{0} \to \bar{D}^{0} (\rho^{0} \to) \pi^{+}\pi^{-}$ & $\frac{1}{\sqrt{2}}(E-C)$ &&
          $0.11^{+0.01+0.02+0+0.01}_{-0.01-0.02-0-0.01}$ & $0.013^{+0.009}_{-0.008}$ \\
        
   \hline \hline
\end{tabular}
\end{center}
\end{table}

\begin{table}[tbhp]
\caption{The same as table \ref{Drho}, but for 
the quasi-two-body decays (top)
$\bar B_{(s)} \to D (K^{*} \to) K \pi$ ($\times 10^{-4}$), 
and (bottom) $\bar B_{(s)} \to \bar D (K^{*} \to) K \pi$ ($\times 10^{-6}$).}
\vspace{-0.5cm}
 \label{DKstar}
\begin{center}
\begin{tabular}{ccccc}
 \hline \hline
Decay Modes    &  Amplitudes &  Data &${\mathcal B}_{\rm {FAT}} $       
&  ${\mathcal B}_{\rm {PQCD}} $   \\
\hline 
   $\bar B \to D(K^* \to) K \pi $ & $V_{cb} V_{ud}^*$ &   & &  \\
   
 $\bar{B}^0 \to D_s^+(K^{*-}\to)K^- \pi^0$ &~~~$ E $~~~ &&
                   $0.11^{+0.02+0+0.02}_{-0.01-0-0.02}$ & 
                   $0.52^{+0.14+0.05+0.05}_{-0.12-0.08-0.00}$\\

 $\bar{B}_s^0 \to D^0(K^{*0}\to)K^+ \pi^-$ &~~~$ C $~~~ & $2.86 \pm 0.44 $&
                      $3.74^{+0.16+0.79+0.04}_{-0.15-0.71-0.04}$ & 
                     $2.86^{+1.67+0.43+0.05}_{-1.33-0.56-0.08}$\\
  
      & $V_{cb} V_{us}^*$  &  & \\
 $B^- \to D^0 (K^{*-}\to) K^- \pi^0$  &~~~$T+C$~~  & &
          $2.04^{+0.04+0.35+0.17}_{-0.05-0.33-0.16}$ & 
          $~~1.67^{+0.71+0.32+0.07}_{-0.53-0.34-0.07}$ \\
 
  $\bar{B}^0 \to D^+(K^{*-}\to)K^- \pi^0$   &~~~$ T $~~~  & &
                   $1.31^{+0+0.28+0.13}_{-0-0.25-0.13}$ & 
                   $1.24^{+0.55+0.15+0.06}_{-0.40-0.18-0.05}$\\
 
  $\bar{B}^0 \to D^0 (\bar K^{*0}\to)K^- \pi^+$ &~~~$ C $~~~ & $0.32 \pm 0.05$ &
                    $0.27^{+0.01+0.06+0.01}_{-0.01-0.05-0.01}$ & 
                    $0.17^{+0.10+0.03+0.00}_{-0.08-0.03-0.01}$\\

  $\bar{B}_s^0 \to D_s^+ (K^{*-}\to)K^- \pi^0$ &~~~$ T+E $~~~ &&
                      $1.42^{+0.01+0.31+0.15}_{-0.01-0.28-0.14}$ & 
                      $1.11^{+0.45+0.20+0.05}_{-0.33-0.21-0.04}$\\              
 \hline
 
  $\bar B \to \bar D (K^* \to) K \pi $ & $V_{ub} V_{cs}^*$ &  & & \\    
  
  $B^{-} \to \bar {D}^{0} (K^{*-} \to ) K^{-}\pi^{0}$  & $C+A$ &&
             $4.46 ^{+0.18+0.74+0.06+0.39}_{-0.18-0.68-0.06-0.39} $ & 
             $1.00^{+0.43+0.20+0.00}_{-0.48-0.27-0.07}$ \\
         
   $B^{-}\to D^{-}(\bar{K}^{*0}\to) K^{-}\pi^{+}$  & A &&
             $0.72^{+0+0+0.03+0.06}_{-0-0-0.01-0.06}$ & 
             $0.21^{+0.10+0.03+0.04}_{-0.06-0.02-0.00}$ \\
    
   $\bar{B}^{0}\to \bar{D}^{0} (\bar K^{*0}\to ) K^{-}\pi^{+}$ & $C$ &&
                $3.48^{+0.15+0.73+0.04+0.31}_{-0.14-0.66-0.04-0.31} $ & 
                $1.96^{+1.01+0.52+0.11}_{-0.87-0.41-0.12} $ \\
    
   $\bar{B}_{s}^{0} \to D_{s}^{-} (K^{*+} \to ) K^{+}\pi^{0}$ & $T+E$  & &
                 $8.59^{+0.14+1.92+0.86+0.76}_{-0.08-1.72-0.82-0.76}$ & 
                 $13.3^{+6.84+0.76+0.80}_{-3.04-0.73-0.79}$     \\
                         
  & $V_{ub} V_{cd}^*$ &  &  &\\      
        
   $B^{-}\to D_{s}^{-} (K^{*0}\to) K^{+}\pi^{-}$  & $ A $ &&
              $0.037^{+0+0+0.002+0.003}_{-0-0-0.001-0.003}$  & 
              $0.014^{+0.008+0.004+0.002}_{-0.003-0.008-0.0002}$    \\
              
   $\bar{B}^{0}\to D_{s}^{-} (K^{*+}\to) K^{+}\pi^{0}$  & $ E$ &&
                $0.005^{+0.0009+0+0.0007+0.0004}_{-0.0004-0-0.0007+0.0004}$  &  
                $0.005^{+0.003+ 0.001+ 0}_ {-0.003- 0.001- 0}$ \\
    
  $\bar{B}_{s}^{0}\to D^{-} (K^{*+}\to ) K^{+}\pi^{0}$ & $T$ &&
                $0.35^{+0 + 0.07 +0.004+0.03}_{-0 - 0.07 -0.004+0.03}$ & 
                 $0.6^{+0.30+0.03+0.04}_{-0.15-0.04-0.04}$    \\

  $\bar{B}_{s}^{0}\to \bar{D}^{0} (K^{*0}\to) K^{+}\pi^{-}$ & $C$  &&
                  $0.16^{+0.01+ 0.03 +0.002+0.01}_{-0.01- 0.03 - 0.002-0.01}$ &  
                  $0.08^{+0.05+0.02+0.00}_{-0.03-0.02-0.00}$    \\
            
   \hline \hline
\end{tabular}
\end{center}
\end{table}

\begin{table}[!hb]
\caption{The same as table \ref{Drho}, but for 
the quasi-two-body decays (top)
$\bar B_{(s)} \to D (\phi\to) K K$ ($\times 10^{-4}$), 
and (bottom) $\bar B_{(s)} \to \bar D (\phi \to) K K $ ($\times 10^{-6}$).}
 \label{Dphi}
\begin{center}
\begin{tabular}{cccc}
 \hline \hline
Decay Modes    &  Amplitudes & ${\mathcal B}_{\rm {FAT}} $       
&  ${\mathcal B}_{\rm {PQCD}} $   \\
\hline    
  $\bar B \to D(\phi \to) K K $ & $V_{cb} V_{us}^*$ &   &  \\
                        
  $\bar{B}_s^0 \to D^0 (\phi \to) K^+ K^-$  &$ C $  
             &$0.193^{+0.008+0.041+0.002}_{-0.008-0.037-0.002}$& \\                                
 
        $ ~~ \to D^0 (\phi \to) K^0 \bar K^0$  &  &
           $0.134^{+0.006+0.028+0.001}_{-0.006-0.026-0.001}$ & 
          \\     
   \hline    
    $\bar B \to D(\phi \to) K K $ & $V_{ub} V_{cs}^*$ &   &  \\
    
   ${B}^{-} \to D_{s}^{-}(\phi \to )K^{+} K^{-}$    & A
                 &$0.75^{+0+0+0.32+0.07}_{-0-0-0.32-0.07}$  & 
                 $0.15^{+0.02+0.01+0.01}_{-0.02-0.01-0.01}$
                  \\
                  
   $ ~~ ~~~~~\to D_{s}^{-}(\phi \to ) \to{K}^{0}{K}^{0}$  &
                &$0.52^{+0+0+0.32+0.05}_{-0-0-0.34-0.05}$ & 
                $0.10^{+0.01+0.01+0.01}_{-0.01-0.01-0.01}$   
                \\                       
   $\bar {B}_{s}^{0} \to \bar{D}^{0}(\phi \to) K^{+} K^{-}$  & C
                           &$2.85^{+0.12+0.60+0.03+0.25}_{-0.12-0.54-0.03-0.25}$  &     \\
                           
   $ ~~~ \to \bar{D}^{0} (\phi \to) \bar{K}^{0} K^{0}$&
                         &$1.98 ^{+0.08+0.42+0.02+0.17}_{-0.08-0.38-0.02-0.17}$   &     \\                                                 
 \hline \hline
\end{tabular}
\end{center}
\end{table}

\begin{table}[tbhp]
\caption{Comparison of results from FAT (${\mathcal B}^v_{ \rm {FAT}} $) 
and PQCD (${\mathcal B}^v_{\rm {PQCD}} $) approach for the virtual effects 
of $B_{(s)} \to D( \rho, \omega \to)K \bar K$ decays, happened when the 
pole masses of $\rho, \omega $ are smaller than the invariant mass of $K \bar K$.}
 \label{Domega}
\begin{center}
\begin{tabular}{cccc}
 \hline \hline
{Decay Modes}     &  ${\mathcal B}^v_{ \rm {FAT}} $   &  ${\mathcal B}^v_{\rm {PQCD}} $  \\
\hline
 $\bar B \to D(\rho \to) K K $ & &   &  \\
 
 $B^-\to D^0 (\rho^-\to) K^- K^0$\;  &
        $7.01^{+0.13+1.26+0.63}_{-0.14-1.16-0.60}\times10^{-5}$ &
        $11.8^{+6.2+0.9+0.7}_{-4.0-1.2-0.9}\times10^{-5}$
        \\
 $\bar B^0\to D^+ (\rho^-\to) K^- K^0$\;  &
            $4.64^{+0.03+1.00+0.49}_{-0.02-0.90-0.47}\times10^{-5}$ &
            $7.93^{+5.01+0.32+0.65}_{-2.93-0.30-0.63}\times10^{-5}$ 
         \\ 
 $\bar B^0\to D^0 (\rho^0\to) K^+ K^-$\;    &
             $1.34^{+0.07+0.28+0.02}_{-0.07-0.26-0.02}\times10^{-6}$ &
             $1.07^{+0.46+0.80+0.01}_{-0.37-0.58-0.01}\times10^{-6}$ 
          \\   
 $\bar B_s^0\to D_s^+ (\rho^-\to) K^- K^0$\; &
            $5.63^{+0+1.18+0.60}_{-0-1.07-0.60}\times10^{-5}$  &
            $6.06^{+3.47+0.04+0.47}_{-2.06-0.04-0.45}\times10^{-5}$ 
          \\   
 $\bar B_s^0\to D^+ (\rho^-\to) K^- K^0$\;  &
       $9.46^{+1.64+0+2.71}_{-0.77-0-1.89} \times10^{-9}$ &
       $4.22^{+0.58+0.90+0.40}_{-0.67-0.65-0.30}\times10^{-8}$ 
          \\   
$\bar B_s^0\to D^0 (\rho^0\to) K^+ K^-$\;  &
        $0.48^{+0.08+0+0.11}_{-0.04-0-0.07}\times10^{-8}$ &
        $1.05^{+0.15+0.23+0.10}_{-0.17-0.15-0.07}\times10^{-8}$ 
          \\   
 \hline
  $\bar B \to \bar D(\rho \to) K K $ & &   &  \\
     $ B^{-} \to D^{-}(\rho^{0} \to)  K^{+}K^{-}$   &                                       
               $1.89^{+0+0.53+0.03+0.04}_{-0-0.46-0.03-0.04} \times10^{-9}$    & 
                $3.22^{+0.52+0.86+0.01}_{-0.45-0.43-0.01}\times10^{-9}$
                 \\
     $ B^{-}\to \bar {D}^{0}(\rho^{-}\to) K^{-}K^{0}$     &                                          
                   $2.51^{+0.10+0.37+0.04+0.17}_{-0.11-0.34-0.04-0.17} \times10^{-9}$   &
                    $0.53^{+0.12+0.25+0.03}_{-0.06-0.17-0.01}\times10^{-9}$  
                    \\                
     $ B^{-}\to D_{s}^{-}(\rho^{0}\to) K^{+}K^{-}$ &                                  
                $8.74^{+0+1.83+0.87+0.77}_{-0-1.66-0.83-0.77}\times10^{-8}$    & 
                $6.26^{+1.69+2.69+0.03}_{-1.30-0.92-0.02}\times10^{-8}$ 
                 \\    
    $\bar{B}^{0} \to D^{-} (\rho^{+}\to) K^{+}K^{0}$    &                                  
               $5.37^{+0.07+1.18+0.06+0.05}_{-0.04-1.07-0.06-0.05} \times10^{-9}$       & 
               $6.87^{+2.05+3.30+0.08}_{1.60-1.01-0.08}\times10^{-9}$
                \\               
     $\bar {B}^{0} \to \bar{D}^{0} (\rho^{0}\to) K^{+}K^{-}$    &           
               $5.73 ^{+0.32+1.26+0.06+0.51}_{-0.31-1.13-0.06-0.51} \times10^{-10}$  & 
               $0.78^{+0.20+0.46+0.08}_{-0.13-0.29-0.06} \times10^{-10}$
                \\
     $\bar {B}^{0} \to D_{s}^{-} (\rho^{+} \to) K^{+}K^{0}$  &                       
              $1.51^{+0+0.32+0.15+0.13}_{-0-0.29-0.14-0.13} \times10^{-7}$  & 
             $2.32^{+0.63+1.00+0.01}_{-0.48-0.34-0.01} \times10^{-7}$ 
               \\
   $\bar{B}_{s}^{0}\to D^{-}(\rho^{+}\to) K^{+}K^{0}$ &                                
               $0.99^{+0.17+0+0.11+0.09}_{-0.08-0-0.11-0.09} \times10^{-9} $  & 
               $7.47^{+1.49+2.42+0.40}_{-0.32-1.83-0.37} \times10^{-9} $ 
               \\
     $\bar{B}_{s}^{0} \to \bar {D}^{0}(\rho^{0}\to) K^{+}K^{-}$              
                 &$0.51^{+0.09+0+0.06+0.04}_{-0.04-0-0.05-0.04}\times10^{-9}$   &
                 $1.85^{+0.36+0.61+0.09}_{-0.32-0.45-0.08} \times10^{-9} $ 
                \\
  \hline 
   $\bar B \to D(\omega \to) K K $ & &   &  \\ 
   
 $\bar B^0\to D^0 (\omega \to) K^+ K^-$\;  & 
       $1.83^{+0.09+0.15+0.03}_{-0.08-0.32-0.03} \times10^{-6}$ &
          \\    
  $\bar B_s^0\to D^0 (\omega \to) K^+ K^-$\;  & 
       $4.01^{+0.70+0+0.92}_{-0.33-0-0.57} \times10^{-9}$ &
          \\         
  $\bar B \to \bar D(\omega \to) K K $ & &   &  \\    
  
          $ B^{-}\to  D_{s}^{-}(\omega \to) K^{+}K^{-}$                         
               &$6.90^{+0+1.45+0.68+0.61}_{-0-1.31-0.65-0.61}\times10^{-8}$   &  
               \\
   $B^{-} \to D^{-}(\omega \to) K^{+}K^{-}$                                 
                &$4.15^{+0+0.68+0.06+0.37}_{-0-0.63-0.05-0.37}\times10^{-9}$  & 
                 \\    
      
     $\bar{B}^{0} \to \bar{D}^{0} (\omega \to) K^{+}K^{-}$     
                  &$6.14^{+0.30+1.16+0.09+0.54}_{-0.28-1.05-0.09-0.54}\times10^{-10}$  &
                   \\   
     $\bar {B}_{s}^{0}\to \bar {D}^{0} (\omega \to) K^{+}K^{-}$  
                  &$4.21^{+0.73+0+0.49+0.37}_{-0.34-0-0.45-0.37}\times10^{-10}$     &  \\
\hline\hline
\end{tabular}
\end{center}
\end{table}

 
 \subsection{Hierarchies of branching fractions}

The decay modes are classified by CKM matrix elements involved, 
Cabibbo favored $V_{cb} V^*_{ud}$, Cabibbo suppressed $V_{cb} V^*_{us}$, 
and doubly Cabibbo suppressed $V_{ub} V^*_{ud}$ and $V_{ub} V^*_{us}$, 
shown in the second column of Tables \ref{Drho},  \ref{DKstar} and \ref{Dphi}.
The hierarchies of branching fractions can be seen clearly from this classification.
The Cabibbo favored decay modes are about two orders larger than the doubly 
Cabibbo suppressed ones of the same type in the same table. 
As a result, these Cabibbo favored decay modes are able to be measured firstly 
by experiments, such as the first four modes of $\bar B_{(s)} \to D (\rho\to) \pi \pi$ 
~\cite{Workman:2022ynf} in table \ref{Drho} and 
$\bar{B}_s^0 \to D^0(K^{*0}\to)K^+ \pi^-$ \cite{LHCb:2014ioa} in 
table \ref{DKstar}. Our results and the experimental data are consistent
within errors.
 
Besides CKM matrix elements, the hierarchy of branching fraction is 
also dependent on contributions from different topological diagrams. 
Similar to the dynamics of two-body hadronic $B$ decays, the color 
favored emission diagram ($T$) is absolutely dominating in the 
quasi-two-body decays. For instance, the topology $T$ dominated 
decay modes, $B^- \to D^0 (K^{*-}\to) K^- \pi^0$, 
$\bar{B}^0 \to D^+(K^{*-}\to)K^- \pi^0$ and 
$\bar{B}_s^0 \to D_s^+ (K^{*-}\to)K^- \pi^0$
 happening through Cabibbo suppressed $V_{cb} V^*_{us}$, are 
 the same order as the Cabibbo favored but only $C$ contributed 
 decay, $\bar{B}_s^0 \to D^0(K^{*0}\to)K^+ \pi^-$. Besides the mode 
 $\bar{B}_s^0 \to D^0(K^{*0}\to)K^+ \pi^-$ with large branching ration
has been measured by LHCb experiment through Dalitz plot analysis 
and isobar model \cite{LHCb:2014ioa}, another mode 
$\bar{B}^0 \to D^0 (\bar K^{*0}\to)K^- \pi^+$ with one order smaller 
branching ration is also measured by LHCb \cite{LHCb:2015tsv}.
The remaining three modes with comparable branching ratio as 
$\bar{B}_s^0 \to D^0(K^{*0}\to)K^+ \pi^-$ are also measurable in 
LHCb and Belle II. Our results of other decays, especially those with 
branching ratios in the range $10^{-6}-10^{-4}$ in 
Tables \ref{Drho},  \ref{DKstar} and \ref{Dphi} are expected to 
be observed in future experiments.
 
 \subsection{Comparison with the results in the PQCD approach}
Since most quasi-two-body decays $ B_{(s)} \to D_{(s)}  (V\to ) P_1 P_2$ 
have not been measured by experiments until now, we list the results 
calculated in PQCD approach  
\cite{Ma:2016csn,Ma:2020dvr,Zou:2022xrr, Fang:2023dcy,Wang:2024enc} 
in the last column of the tables for comparison. As we have stated 
in Sec. \ref{Introduction}, LCDA of P-wave $P_1 P_2$ meson pair 
from resonance is described by RBW model in PQCD, which is the 
same theme adopted in the FAT approach for $V \to P_1 P_2$. 
The two approaches are effectively compatible for intermediate 
resonance strong decays, the main difference between them 
is the calculation of the weak decays of $B$ to $D$ meson 
and a vector resonance.

As known, $T$ diagram is proved to be factorizable at 
all orders of $\alpha_s$ for these decays, thus the perturbative 
calculation is reliable. Our results of the $T$ diagram 
dominating decay modes in Tables \ref{Drho} and \ref{DKstar} 
are in good agreement with PQCD's predictions. The magnitude 
of topologies $C$ is larger than $E$, $\left|C \right| > \left|E\right|$,  
in the FAT approach as shown in eq.(\ref{fitpara}), while $C$ is 
approximately equal to $E$, $\left|C \right| \sim\left|E\right|$,
in the PQCD approach \cite{Li:2008ts}, because it is sensitive to the 
power corrections and high order contributions which are hard 
to be calculated in PQCD approach. Therefore, it is easy to find 
in Tables \ref{Drho} and \ref{DKstar} that the results of the FAT 
approach for the decays dominated only by $C$, larger than those 
in the PQCD approach, are in better agreement with the current 
experimental data. However, our results of decay modes with 
only power suppressed $E$ contribution are a little smaller than 
those in PQCD, which need to be tested by the future experiments. 
At last, we emphasize that the branching ratios of decays in the FAT 
approach are more precise than those in the PQCD in 
Tables \ref{Drho}, \ref{DKstar} and \ref{Domega}.
The reason is that the topological amplitudes in the FAT including the 
nonfactorizable QCD contributions were extracted through a global 
fit with experimental data of these decays, while large 
uncertainties arise from non-perturbative parameters and QCD 
power and radiative corrections in the PQCD.

 \subsection{The virtual effects of $B_{(s)} \to D( \rho, \omega \to)K \bar K$ }
 
Contrary to the quasi-two-body decays through 
$\rho \to \pi \pi$, $K^* \to K \pi$ and $\phi \to K K$ proceeding by 
the pole mass dynamics, i.e., the pole mass is larger than the 
invariant mass threshold of two final states, the other modes 
with strong decays by $\rho, \omega \to KK $ can only happen 
by $\rho, \omega$ off-shell effect. It is also called the Breit-Wigner 
tail (BWT) effect, which has also appeared in charmed quasi-two-body 
decay with off-shell $D^*$ resonance \cite{Zhou:2021yys,Chai:2021kie}
and charmless one through $\rho, \omega$ resonances 
\cite{Wang:2016rlo, Li:2016tpn, Li:2018psm, Fan:2020gvr,Zhou:2023lbc}.
We denote branching ratios of this kind of decays by $\mathcal{B}^v$ 
and their numerical results are listed on Table \ref{Domega},
together with the PQCD's predictions for $B \to D (\rho \to) K  K $ 
in last column.

Apparently, the branching ratios of $\mathcal{B}^v$ modes are  
approximately two orders smaller than those of ${\cal B}$ modes 
in Table \ref{Drho}, that is, the BWT effect in $B \to D (\rho \to) K  K$ 
is only about $1 \% $ of the on-shell resonance contribution, 
$B \to D (\rho \to) \pi \pi$. In Tables \ref{Dphi} and \ref{Domega}, 
one can see that all the intermediate states of $\rho, \omega $, 
and $\phi$ can decay into $K  K$ via virtual effects (for $\rho, \omega $) 
or pole mass dynamics (for $\phi $). However, different from 
neutral states $\rho^0, \omega, \phi$, the charged $\rho^{\pm}$ 
is the unique resonance contributing to the charged meson pair 
$K^{\pm} K^0$ in the low mass region of $K^{\pm} K^0$ system, 
which have been measured recently by Belle II collaboration 
\cite{Belle-II:2023gye} based on a study of the small $m(K^- K_S^0)$ 
invariant mass for $B^- \to D^0 K^- K_S^0$ and 
$\bar B^0 \to D^+ K^- K_S^0$. With $\rho^-$-like resonances 
and non-resonance contribution, 
the branching ratios are 
$\mathcal{B} (B^- \to D^0 K^- K_S^0)=
(1.89\pm 0.16\pm0.10) \times 10^{-4}$ 
and 
$\mathcal{B} (\bar B^0 \to D^+ K^- K_S^0)=
(0.85\pm 0.11\pm0.05) \times 10^{-4}$,
respectively. 
Our result for only ground state $\rho (770)$ is
$\mathcal{B} (B^- \to D^0 (\rho^- \to) K^- K^0)=
(7.01^{+0.13+1.26+0.63}_{-0.14-1.16-0.60})\times10^{-5}$ 
and
$\mathcal{B} (B^- \to D^0 (\rho^- \to) K^- K^0)=
(4.64^{+0.03+1.00+0.49}_{-0.02-0.90-0.47})\times10^{-5}$ 
in Table \ref{Domega}, which can reach a proportion of about 
$20 \%$ of above measured all $\rho^-$ resonant and 
non-resonant components (considering half of branching ratios 
of $K^0$ or $\bar K^0$ to become $K_S $).
The similar mode $\bar B_s^0\to D_s^+ (\rho^-\to) K^- K^0$ 
with comparable branching ratio 
$(5.63^{+0+1.18+0.60}_{-0-1.07-0.60})\times10^{-5}$
is suggested to be measured in LHCb and Belle II.

 The study of invariant mass of neutral $K^+ K^-$ system for 
 $\bar B \to D K^+ K^- $ in experiments is relatively complex, 
 as it involves various resonances $\rho^0, \omega$ and $\phi$ 
 as well as non-resonances. Especially, in the low-mass region 
 of $m(K^+ K^-)$, the BWT effects from neutral resonance 
 $\rho^0$ and $\omega$ in decay modes such as 
 $\bar B^0\to D^0 (\rho^0 \to) K^+ K^-$ and 
 $\bar B^0\to D^0 (\omega \to) K^+ K^-$, 
 $\bar {B}_{s}^{0}\to D^{0} (\rho^0 \to) K^{+}K^{-}$ and 
$\bar {B}_{s}^{0}\to D^{0} (\omega \to) K^{+}K^{-}$, 
 are pretty much the same in Table \ref{Domega}, 
even though the decay widths of $\rho$ and $\omega$ meson 
are very different, shown in Table \ref{tab:mass and width}.
As we have mentioned in \cite{Zhou:2021yys,Zhou:2023lbc}, 
the BWT effects in these decays are not very sensitive to the 
widths of resonances. It can be attribute to the behavior of the 
Breit-Wigner propagator in eq.(\ref{RBW}) describing off-shell 
resonance, where the invariant mass square $s$ is far away 
from the on-shell mass of resonance, e.g. the real part,  
$\mid s-m_{\rho, \omega}\mid$, of denominators of 
Breit-Wigner formula is much larger than the imaginary 
part $ i\, m_{\rho, \omega}\, \Gamma_{\rho, \omega}$.

Finally, the comparison of BWT effect in $B \to D (\rho \to) K  K$ 
between the FAT and PQCD approaches is very similar with that 
of the on-shell resonance contributions in $B \to D (\rho \to) \pi \pi$. 
They are in agreement for $T$ diagram dominated modes, 
but different from those dominated by $C$ and $E$ diagrams. It 
indicates again that no matter for on-shell resonance or for off-shell 
one, the mechanisms or models applied by the two approaches are
effectively consistent.
 
\section{Conclusion}\label{sec:4}
Motived by the measurements of three-body charmed $B$ meson 
decays with resonance contributions, especially ground state 
resonance contributions,  from Babar, LHCb and Belle (II), we 
systematically analyze the corresponding quasi-two-body 
decays $B_{(s)} \to D_{(s)} (V \to)  P_1 P_2 $ through intermediate 
ground states $\rho, K^*, \omega$ and $\phi$. They proceed 
by $b \rightarrow \,c\, $ or $b \rightarrow \,u\, $ transitions to 
a $D_{(s)} V$ intermediate state with $V$ as a resonant 
state which decays consequently into final states $P_1, P_2$ via 
strong interaction. We utilize the decay amplitudes extracted 
from the two-body charmed $B$ decays in the FAT approach for 
the first subprocess $B_{(s)} \to D_{(s)} V$ and RBW function 
for the narrow widths resonances $V$ as usually done in 
experiments and the PQCD approach. We categorize 
$B_{(s)} \to D_{(s)} (V \to)  P_1 P_2 $ into four groups according to 
different vector resonance, $B_{(s)} \to D_{(s)} (\rho \to)  \pi \pi $, 
$B_{(s)} \to D_{(s)} (K^* \to) K \pi $, $B_{(s)} \to D_{(s)} (\phi \to) K K $ 
and $B_{(s)} \to D_{(s)} (\rho, \omega \to) K K $, where the former three
kinds of modes decay by pole dynamics, and the last one by BWT effect. 

We calculate the branching ratios of all the four kinds of 
decay modes in the FAT method. Our results are consistent 
with the data by Babar, LHCb and Belle (II). Our predictions of 
order $10^{-6}-10^{-4}$ without any experimental data are 
hopeful to be observed in the future experiments. The FAT 
approach and the PQCD approach have effectively compatible 
mechanism of resonant state strong decays. Meanwhile, 
their treatments on the weak decays of $B$  to a $D$ meson 
and a vector resonance are different. Since the calculation 
of the first subprocess is done by a global fit with experimental 
data in the FAT approach, our results for the color suppressed 
diagram dominating modes are larger than those in the PQCD 
approach whose information on nonperturbative contribution 
and $m_c/m_b$ power corrections are not included so far. 
In addition, our results have significantly less theoretical 
uncertainties  due to accurate nonfactorizable parameters 
extracted from experimental data. 

The fourth type of modes happen through the tail effects of 
$\rho$ and $\omega$ resonance to $KK$. It's found that 
the BWT effect of $\rho$ resonance approximately is about 
two orders smaller than $\rho$ on-shell resonance contribution, 
which induce that they are usually ignored in experimental 
analysis. However charged $\rho^{\pm}$-like resonance of 
low mass region of $K^{\pm} K^0$ system have been started 
to be studied in Belle II recently. The comparable modes, 
such as $\bar B_s^0\to D_s^+ (\rho^-\to) K^- K^0$, also 
have the potential to be measurabled in LHCb and Belle II.

\section*{Acknowledgments}
The work is supported by the National Natural Science Foundation of China
under Grants No.12075126 and No.12105148.

\bibliographystyle{bibstyle}
\bibliography{refs}

\end{document}